# Free Control of Far-Field Scattering Angle of Transmission Terahertz Wave using Multilayer Split-Ring Resonators' Metasurfaces

Ying Tian, Xufeng Jing, Haiyong Gan, Chenxia Li, Zhi Hong

*Abstract*—To enhance transmission efficiency of Pancharatnam–Berry (PB) phase metasurfaces, multilayer split-ring resonators were proposed to develop encoding sequences. As per the generalized Snell's law, the deflection angle of the PB phase encoding metasurfaces depends on the metasurface period's size. Therefore, it is impossible to design an infinitesimal metasurface unit; consequently, the continuous transmission scattering angle cannot be obtained. In digital signal processing, this study introduces the Fourier convolution principle on encoding metasurface sequences to freely control the transmitted scattering angles. Both addition and subtraction operations between two different encoding sequences were then performed to achieve the continuous variation of the scattering angle. Furthermore, we established that the Fourier convolution principle can be applied to the checkerboard coded metasurfaces.

## I. INTRODUCTION

METASURFACE is a 2D metamaterial that can produce phase mutations in the subwavelength propagation distance range [1-5]. A metasurface is arranged either periodically or non-periodically as per the unit structure, it can thus effectively control the amplitude, phase, polarization, and propagation mode of the incident electromagnetic waves [6-10]. A metasurface can then realize various functional devices such as holographic phase plate, vortex optical phase plate, flat lens, perfect absorption, invisibility cloak, and polarization conversion [11-15].

Recently, the concept of coded metasurfaces that can connect physical space and digital space has been proposed [16]. Cui et al. proposed using digital coding to describe the structure of metasurface units and adjust electromagnetic waves by changing the spatial arrangement of coding units. Nevertheless, the coding unit structure has a certain size; when the unit structure is arranged, the period of the metasurface cannot be continuously changed [17]. As per the generalized Snell's law, the electromagnetic beam's deflective angle depends on the metasurface period [18]. Therefore, the metasurface based on the arrangement of the discrete coding unit structure cannot realize a free control and continuous change of the electromagnetic wave scattering angle. To obtain the continuously changing refraction angles, we introduced the convolution calculation principle in digital signal processing as well as performed addition and subtraction operations on differently coded metasurface sequences.

From the phase mutation mechanism perspective, metasurfaces can be divided into two categories: geometric phase and resonant phase metasurfaces [19]. The phase mutation of the resonant phase metasurface is attributed to structural resonance, resulting in a narrower working bandwidth and a reduced tolerance of device characteristics. The geometric phase electromagnetic metasurface is a metasurface composed of the same artificial microstructure using different rotation angles. By simply changing the rotation angle of the microstructure, it can realize the phase mutation of the reflected or transmitted waves. Such changes help in realizing the manual control of the phase gradient or the distribution, thereby considerably reducing the complexity of designing and processing of the metasurface. However, for a transmitted geometric phase metasurface, the maximum efficiency of metasurfaces was theoretically obtained to be ~25% for an individual structure [20]. Recently, several bilayer metasurfaces were proposed such as double split loop resonators [21], complementary split-ring resonators, and enclosed H-shaped structure [22]. However, most transmitted bilayer metasurfaces suffered from low transmission efficiencies.

In this study, we proposed three-layer split-ring resonators to develop encoding metasurfaces using different sequences. The complete phase control of metasurfaces with higher transmission can be achieved by simply rotating the oscillator unit structure. In a three-layer structure, the efficiency of metasurfaces can then be improved based on the electromagnetic coupling between layers. The Fourier convolution principle in digital signal processing is then applied to different sequences of coded metasurface to obtain the coding sequence. Thus, we freely controlled the scattering angle of the electromagnetic beam using the addition and subtraction calculations on two different encoding sequences.

Ying Tian, Xufeng Jing, Chenxia Li, and Zhi Hong are with Institute of Optoelectronic Technology, China Jiliang University, Hangzhou 310018, China, (The corresponding author of Xufeng Jing, e-mail: jingxufeng@cjlu.edu.cn).

Haiyong Gan is with National Institute of Metrology, Beijing, China (The corresponding author of Haiyong Gan, e-mail: ganhaiyong@nim.ac.cn).

This work was supported by Natural Science Foundation of Zhejiang Province under Grant No.LY20F050007; National Natural Science Foundation of China under Grant No.61875179. National Key Research and Development Project of China under Grant 2017YFF0206103.

## II. THEORY

### A. Far-field scattering angle of encoding metasurfaces

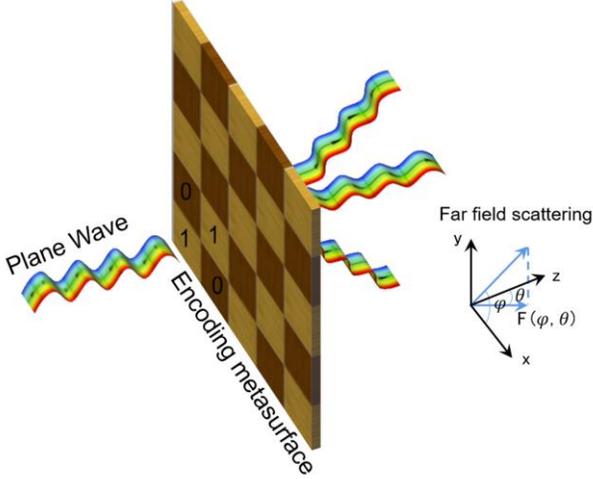

Fig. 1. Schematic of a 1-bit encoding metasurface

Figure 1 shows the schematic of a 1-bit encoding metasurface. The phase difference of these two metasurface atoms is 180 ° as the digital 0 and 1 unit, thus constructing the 1-bit digital coded metasurface. Both $d_x$ and $d_y$ are the sizes of the coding unit structure, whereas $\Gamma_x$ and $\Gamma_y$ are the periods of the coded metasurface. The whole encoding metasurface is composed of M × N units, whereas the corresponding transmission phase is $\varphi(m, n)$ when the encoding unit is in the position (m, n). As per the traditional phased array antenna theory [23], the far field scattering of encoding metasurface can be expressed as follows:

$$F(\theta, \varphi) = \sum_{m=1}^{M} \sum_{n=1}^{N} \exp\{-i\varphi(m,n) - i[k_0 d_x \left(m - \tfrac{1}{2}\right)\sin\theta\cos\varphi + k_0 d_y \left(n - \tfrac{1}{2}\right)\sin\theta\cos\varphi]\}, \quad (1)$$

where $k$ is the wave vector and $\theta$ and $\varphi$ are the angles of elevation and azimuth of scattering, respectively. For the 1-bit encoding metasurface, $\varphi(m,n) = 0$ or $\pi$. Therefore, equation 1 can be simplified as follows:

$$F(\theta, \varphi) = \sum_{m=1}^{M} \exp\{-i[k_0 d_x \left(m - \tfrac{1}{2}\right)\sin\theta\cos\varphi + m\pi]\} \sum_{n=1}^{N}\{-i[k_0 d_y \left(n - \tfrac{1}{2}\right)\sin\theta\cos\varphi + n\pi]\}. \quad (2)$$

By calculating the finite term accumulation in Eq. (2), the intensity of far-field scattering on the encoding metasurface can be expressed as follows:

$$|F(\theta,\varphi)| = MN \operatorname{Sinc}\left(m\pi \left(p + \tfrac{1}{2}\right) - \tfrac{m}{2} k d_x \sin\theta \cos\varphi\right) \operatorname{Sinc}\left(n\pi \left(q + \tfrac{1}{2}\right) - \tfrac{n}{2} k d_y \sin\theta \sin\varphi\right), \quad (3)$$

where p, q = 0, ±1, ±2 .... From Eq.(3), |F(θ,φ)| attains the first extreme value when the following conditions are met;

$$\varphi = \pm \tan^{-1}\tfrac{d_x}{d_y}, \quad \varphi = \pi \pm \tan^{-1}\tfrac{d_x}{d_y} \quad (4)$$

$$\theta = \sin^{-1}\left(\tfrac{\pi}{k}\sqrt{\tfrac{1}{d_x^2} + \tfrac{1}{d_y^2}}\right) \quad (5)$$

When $\Gamma_x = 2d_x$, $\Gamma_y = 2d_y$. Eq.(5) can be simplified further as follows:

$$\theta = \sin^{-1}\left(\lambda\sqrt{\tfrac{1}{\Gamma_x^2} + \tfrac{1}{\Gamma_y^2}}\right) \quad (6)$$

when $\Gamma_x \to \infty$ or $\Gamma_y \to \infty$. Similarly, equation (6) can be simplified further as follows:

$$\theta = \sin^{-1}\left(\tfrac{\lambda}{\Gamma}\right), \quad (7)$$

where Γ is the physical period length of a 1D gradient coded sequence.

### B. Fourier convolution principle of encoding metasurface

Based on Eq. (7), the scattering angle of the encoding metasurface depends on the period of the encoding metasurface. However, the period of the encoding metasurface determines the size of the digital encoding unit structure. Because we cannot design an infinitely small coding metasurface unit structure, the period of the coding metasurface cannot be continuously changed, which indicates that the scattering angle of the coding metasurface cannot be continuously changed. Note that the encoded metasurface sequence has a Fourier transform relationship with its far-field scattering direction. Similarly, the time signal and frequency domain signal in the digital signal processing are a pair of Fourier transform. Therefore, the encoding metasurface sequence can be similar to the time signal in digital signal processing, whereas the corresponding far-field scattering of the coded metasurface can be similar to the frequency domain in the digital signal processing. As per the digital convolution theorem, the mathematical formula can be expressed as follows:

$$f(t) \cdot g(t) \overset{FFT}{\Longleftrightarrow} f(\omega) \ast g(\omega) \quad (8)$$

Furthermore, the time-domain signal in Eq. (8) can be similar to the encoding metasurface sequence, whereas the frequency domain signal in Eq. (8) can be equivalent to the far field scattering of the encoding metasurface. Therefore, Equation (8) can be expressed as follows:

$$f(x_\lambda) \cdot g(x_\lambda) \overset{FFT}{\Longleftrightarrow} f(\sin\theta) \ast g(\sin\theta), \quad (9)$$

where $x_\lambda = \Gamma/\lambda$ is the encoding metasurface sequence and $\theta$ is the far-field scattering angle. In Eq. (8), when $g(\omega) = \delta(\omega - \omega_0)$, the digital convolution theorem can be derived as follows:

$$f(t) \cdot e^{i\omega_0 t} \overset{FFT}{\Longleftrightarrow} f(\omega) \ast \delta(\omega - \omega_0), \quad (10)$$

where $e^{i\omega_0 t}$ is the time-shift item, and its frequency spectrum expression is $\delta(\omega_0)$. Similarly, equation (9) on the encoding metasurface can be expressed as follows:

$$f(x_\lambda) \cdot e^{i\sin\theta_0 x_\lambda}$$
$$\overset{FFT}{\Longleftrightarrow} F(\sin\theta) \ast \delta(\sin\theta - \sin\theta_0) = F(\sin\theta - \sin\theta_0), \quad (11)$$

where $e^{i\sin\theta_0 x_\lambda}$ is an encoding metasurface sequence with gradient phase distribution, and $f(x_\lambda) \cdot e^{i\sin\theta_0 x_\lambda}$ is a new encoding sequence. Equation (11) shows the product of an initial coded metasurface sequence, whereas a coded metasurface sequence with a gradual phase distribution causes the coded metasurface far field scattering pattern to deviate from its initial scattering direction using $sin\theta_0$. The scattering angle of the new coded sequence can then be expressed as follows: $\theta' = sin^{-1}(sin\theta_1 + sin\theta_2)$, where $\theta_1$ and $\theta_2$ are scattering angles using the two initial coding sequences.

### III. DESIGN OF UNIT CELL OF ENCODING METASURFACES

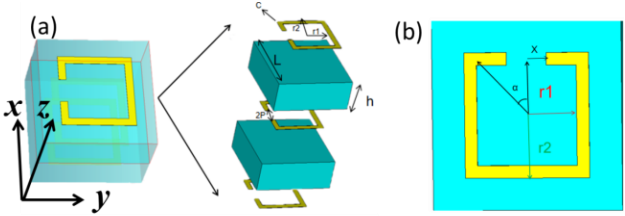

Fig. 2. The element structure of the PB phase encoding metasurface.

To design an efficient coding metasurface, we should design the coded metasurface unit structure. Both the magnetic and the electric resonances should be precisely controlled for the transmissive metasurfaces. For a single-layer metasurface structure, the low efficiency can be predicted because of the only electric response realized in such a structure. In a multilayer system, both electric and magnetic resonances can be achieved. Here, as shown in Fig. 2, we propose the PB geometric phase unit cell with multilayer split-ring resonators. For designing metamaterials, the split-ring resonator structure is the most representative microstructure. This type of a structure is often used to design photonic devices using different functions. The original negative refractive index metamaterials were realized using this type of a split-ring resonance structure. After the precise design, the structural parameters of the unit structure can be determined. In Fig. 2(a), the geometric parameters for three split-ring resonators in each layer having a thickness of 200 nm are similar. The length of unit cell is $L = 76$ μm, whereas the two PI substrates having the same thickness separate the three notched metal rings. The permittivity of PI is $\varepsilon_r = 3.5$ with the loss tangent of $tan\delta = 0.0027$, whereas its thickness is h = 30 μm. In Fig. 2(b), $r_1 = 20$, $r_2 = 25$, and $p = 8$ μm. As per the PB geometric phase theory [12], when the following conditions are met as follows:

$$\begin{cases} |mag(T_{xx})| = |mag(T_{yy})| \\ |arg(T_{xx}) - arg(T_{yy})| = \pi \end{cases}, \quad (12)$$

Note that a left-handed circularly polarized light is perpendicular to the incident light on the coding unit structure. The transmission phase change of the transmitted cross-polarized right-handed circularly polarized light is twice the rotation angle of the encoding unit cell. Both $T_{xx}$ and $T_{yy}$ are the transmission coefficient of the same polarization when a linearly polarized electromagnetic wave enters the periodic structure of a unit particle. Moreover, when two cross-polarized linearly polarized light rays are incident on the periodic structure of the element particle, the designed element structure agrees with the PB phase variation characteristics when the transmission coefficient amplitude of the same polarization is considerable, whereas the phase difference between the two polarizations is π.

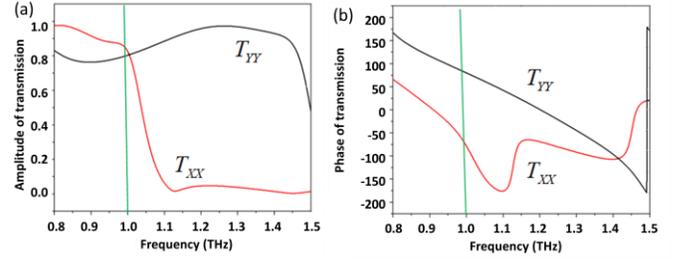

Fig. 3. (a) The co-polarization transmission coefficient and (b) the co-polarization phase difference when the plane wave with *x* polarization or *y* polarization is incident.

Using the finite integral method [24-27], Fig. 3 shows the transmission coefficient and phase difference between the two cross polarization waves for the initial element structure. The transmission amplitude is $|T_{xx}|\approx|T_{yy}|\approx 0.8$, whereas the phase difference between $T_{xx}$ and $T_{yy}$ at 1.0 THz is approximately equal to 180°. In Fig.2(b), $\alpha$ shows the counterclockwise rotation angle of the split-ring resonators. As per the PB phase principle, the phase change of the cross-polarization reaches 0°, $\frac{\pi}{2}$, π, and $\frac{3\pi}{2}$, respectively, when $\alpha = 0°$, $\alpha = 45°$, $\alpha = 90°$, and $\alpha = 135°$. Note that the PB phase characteristic is attributed to the incident circularly polarized wave. For the four rotated units, two-bit coded particles can be constructed as "00," "01," "10," and "11," respectively. Moreover, as shown in Table 1, they can be represented as numeric codes as "0," "1," "2," and "3".

Table 1. Two-bit encoding particles at 1 THz

| Number | 0 | 1 | 2 | 3 |
|---|---|---|---|---|
| $\alpha$ | 0° | 45° | 90° | 135° |
| 2-bit | 00 | 01 | 10 | 11 |
| Phase | −33.04 | 57.02 | 147.1 | 237.6 |

### IV. FOURIER CONVOLUTION OPERATION OF ENCODING METASURFACE SEQUENCES

As per the coding unit particles designed in the previous part, we can develop a two-bit PB phase coded metasurfaces. Firstly, as shown in Fig. 4(a), three basic coding sequences named as S1(01230123…), S2(0011223300112233…), and S3(000111222333…) were designed. Each color line show the basic structure of 1 × 16 with red, blue, yellow, and green, corresponding to the coded particles of 0, 1, 2, and 3, respectively. Fig. 4(b) shows the corresponding structure diagram for these three encoding metasurface sequences. The entire structure comprises 16 × 16 coded particles. Using the finite integral method, the scattering characteristics of the coded metasurface sequences can be accurately calculated. For this simulation, the boundary conditions along x, y, and z axes should be set as open space boundary conditions. The incident light is the left-circularly polarized light at 1 THz; its incident is perpendicular to the −z direction.

In both Fig. 4(c) and Fig. 4(d), the transmitted scattering angles of S1, S2, and S3 are 68.1°, 29°, and 18.7°, respectively.

As per Eq. (7), the transmitted scattering angles of S1, S2, and S3 can be theoretically calculated as 80.6°, 29.5°, and 19.2°. The results of both theoretical calculation and numerical simulation are slightly different for S1, which may be attributed to lesser particles of the encoding metasurface. For both S2 and S3, the theoretical results agree with those of numerical simulations. Note that Fig. 5 shows the 3D scattering patterns at different incident angles for S1, S2, and S3.

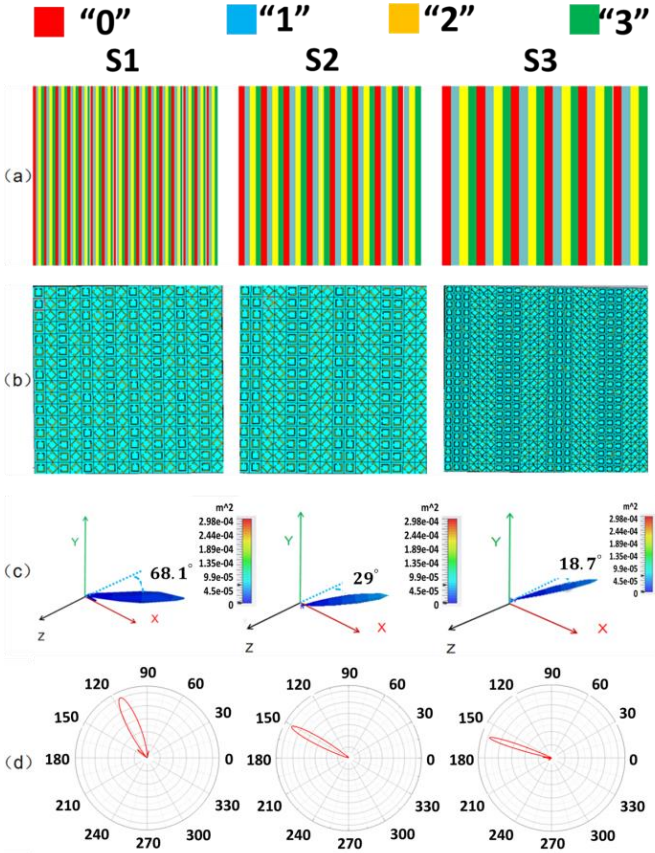

Fig. 4. (a) Schematic of three encoding sequences for S1, S2, and S3, respectively. (b) Corresponding structure diagram of the three coded metasurface sequences. (c) 3D scattering patterns of three sequences, (d) 2D scattering patterns of three sequences.

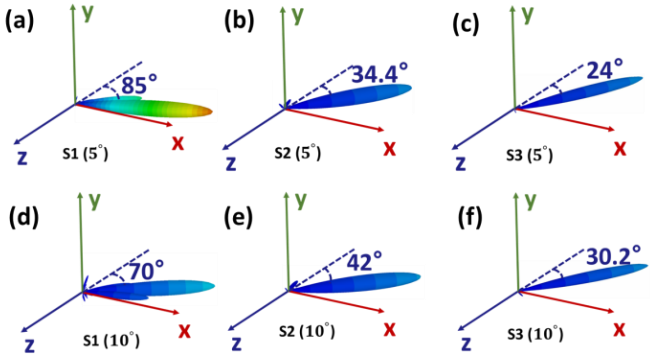

Fig. 5. 3D scattering patterns at different incident angles. (a) S1(5°), (b) S2(5°), (c) S3(5°), (d) S1(10°), (e) S2(10°), and (f) S3(10°).

Based on the scattering patterns of encoding metasurfaces of S1, S2, and S3, only discrete scattering angles can be obtained. However, a continuous variation of scattering angle cannot be achieved. Using these traditional encoding metasurfaces, only limited number of scattering deflected angles can be achieved because the minimum gradient encoding sequence is multiplied using a discrete integer and not a continuous real number [16]. As per the Fourier convolution principle of the encoding metasurface sequences, the transmitted scattering angle can be flexibly controlled using the addition and subtraction operations on the encoding metasurface sequences. Note that, in Eq.(11), the result of multiplying the phases of two encoding metasurfaces is equivalent to evaluating the modulus of their coding digits. Therefore, the four-bit operation on the encoding digits for the encoding metasurface sequences can be implemented as follows: 0 + 0 = 0, 0 + 3 = 3, 2 + 2 = 0, 3 + 2 = 1, and 3 + 3 = 2. Furthermore, in Eq. (11), the subtraction operation on two encoding metasurface sequences of $f(x_\lambda)$ and $e^{i \sin\theta_0 x_\lambda}$ can be implemented; moreover, a new encoding sequence of $f(x_\lambda) \cdot e^{-i \sin\theta_0 x_\lambda}$ can be obtained. In fact, the encoding sequence $e^{-i \sin\theta_0 x_\lambda}$ shows the opposite sequence of $e^{i \sin\theta_0 x_\lambda}$. For example, the basic encoding sequence S1 can be shown as S1(01230123…), whereas the encoding sequence of (-S1) can be expressed as follows: -S1(32103210…). Therefore, Fourier convolution operation of three basic sequences S1, S2, and S3 can be performed as follows: S4 = S1 + S2 = 01302312…, S5 = S1 − S2 = 30011223…, S6 = S2 + S3 = 00123311…, and S7 = S2 − S3 = 3303000…. Here, the subtraction operation of Fourier convolution on the encoding metasurface sequences can be understood as follows: S5 = S1 + (-S2) and S7 = S2 + (-S3). Figure 6 shows both addition and subtraction operation of encoding metasurface sequences of S2 and S3. After performing the Fourier convolution calculation for the basic encoding sequences, the new encoding metasurface sequences of S6 and S7. The scattering angle after the addition and subtraction operations and the scattering angles can be obtained as $\sin\theta_4 = \sin\theta_1 + \sin\theta_2$, $\sin\theta_5 = \sin\theta_1 - \sin\theta_2$, $\sin\theta_6 = \sin\theta_2 + \sin\theta_3$, and $\sin\theta_7 = \sin\theta_2 - \sin\theta_3$.

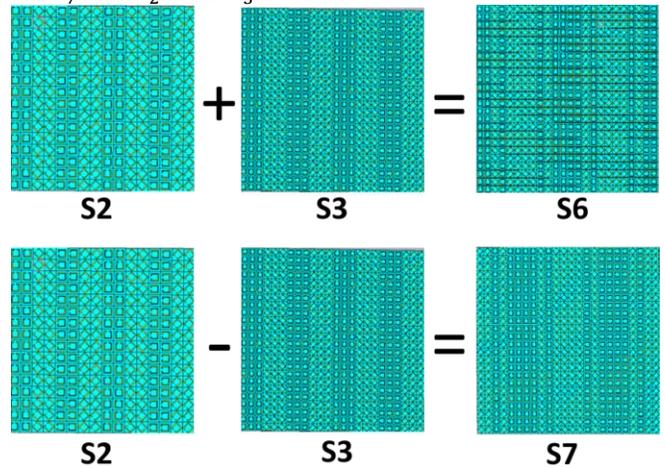

Fig. 6. (a) Addition operation for two basic coding sequences S2 and S3, a new sequence S6 is obtained. (b) Subtraction operation for two basic coding sequences S2 and S3, a new sequence S7 is obtained.

Furthermore, from the following equations, $\sin\theta_4 = (\lambda/4L) + (\lambda/8L)$, $\sin\theta_5 = (\lambda/4L) - (\lambda/8L)$, $\sin\theta_6 =$ $(\lambda/8L) + (\lambda/12L)$, $\sin\theta_7 = (\lambda/8L) - (\lambda/12L)$, and $\sin\theta_4 > 1$, indicating that the sequence S4 cannot realize abnormal transmission. The theoretical results of the scattering angles can be calculated as 29.56 °, 55.32 °, and 9.46 ° corresponding to the encoding metasurface sequences S5, S6, and S7. For the new encoding metasurface sequences of S5, S6 and S7, the scattering patterns are shown in Fig. 7. The transmitted scattering angles of 29.7 °, 54.1 °, and 9.1 °can then be obtained by performing numerical simulations. We thus conclude that the numerical results agree with the theoretical results. Note that 3D scattering patterns at different incident angles for S5, S6 and S7 are shown in Fig. 8.

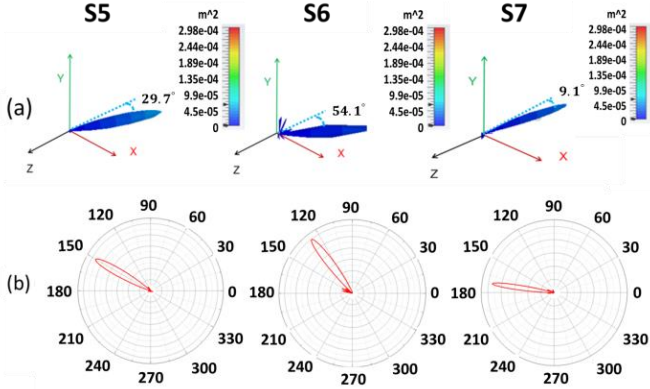

Fig. 7. (a) 3D transmitted scattering pattern and (b) 2D scattering pattern for the encoding metasurface sequences of S5, S6, and S7.

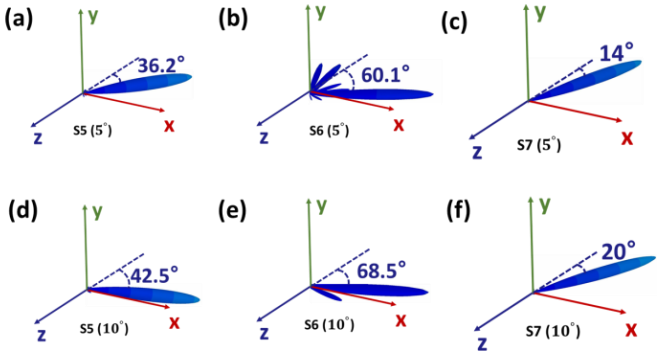

Fig. 8. 3D scattering patterns at different incident angles: (a) S5(5°), (b) S6(5°), (c) S7(5°), (d) S5(10°), (e) S6(10°), and (f) S7(10°).

For the non-basic metasurface sequences of S5, S6, and S7, the Fourier convolution principle, including the addition and the subtraction operations, can be performed. For instance, S8 = S6 + S7 = 33113311…, and S9 = S6 − S7 = 111333111333…. We then confirmed that S8 and S9 belong to the one-bit encoding sequences after addition and subtraction operations. The corresponding encoding periods of sequences S8 and S9 are 4L and 6L, respectively. Fig. 9(a) shows the schematic of the encoding metasurfaces of S8 and S9. Although the encoding period of the two sequences satisfies an integer multiple of L, they do not include encoding unit structures "0" and "2."Thus, they are not part of basic sequences. As per Eq. (7), the scattering angle of S8 can be calculated as follows: $\sin\theta_8 = \lambda/4L > 1$. Fig.9(b) shows the 3D scattering patterns for both S8 and S9. As per the scattering characteristics of S8, the scattered beams are divided into two parts, most of which propagate along the interface of the coded metasurface. The generated beams in Figure 9 show significant sidelobes because of the inadequate number of coded particles that were used. These sidelobes can then be minimized by adding the number of coded metasurface particles. With increase in the number of coding particles, the amount of computation accordingly increases. Therefore, to calculate the scattering properties, we selected a suitable number of coding particles.

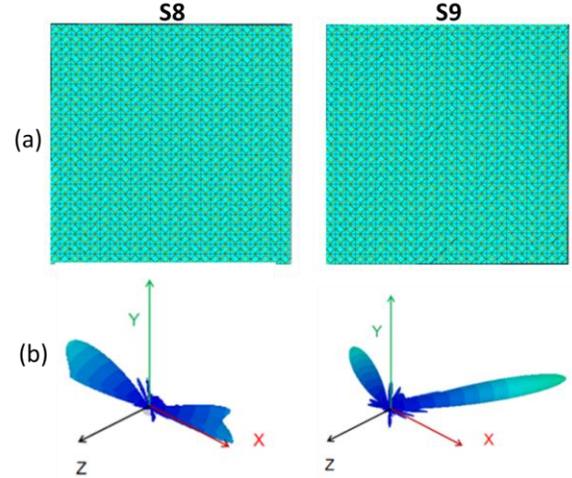

Fig. 9. (a) Encoding metasurface sequences for S8 and S9, respectively. (b) The corresponding 3D transmitted scattering patterns.

To demonstrate the application of Fourier convolution to other coded metasurface sequences, the addition operation on checkerboard coded metasurface sequences is designed. The checkerboard coded pattern can then be expressed as matrix M with $M = \begin{pmatrix} 0 & 2 \\ 2 & 0 \end{pmatrix}$. In matrix M, as shown in Fig. 10(a), each number represents three by three coded particles. Figure 10(b) shows the schematic of the encoding metasurface. Figure 10(c) shows the scattering pattern of this chessboard encoding metasurface. The transmitted scattering angle of 68.5 ° for the chessboard encoding mode can then be calculated as per $\theta = \sin^{-1}(\sqrt{2}\lambda/\Gamma)$.

Then, we can introduce a gradient coding sequence S whose horizontal extension is coded as "00112233…." Figure 11(a) and (b) show the schematic of this gradient encoding metasurface, whereas the corresponding scattering pattern is shown in Fig.11(c). We then performed the Fourier convolution calculation on the checkerboard coded sequence as well as the gradient encoding metasurface sequence. The addition operation between two encoding sequences can then be implemented as follows: $S_{MS} = M + S = 00130033 \dots / 22312211 \dots$. Figs. 12(a) and 12(b) show the new encoding metasurface sequence, and the corresponding transmitted scattering pattern is shown in Fig. 12(c). The whole pattern is tilted away from the z-direction. Note that, although the presented work is completely based on numerical simulations of encoding metasurfaces, it is feasible to prepare such multilayer encoding metasurfaces [11-13, 28-35].

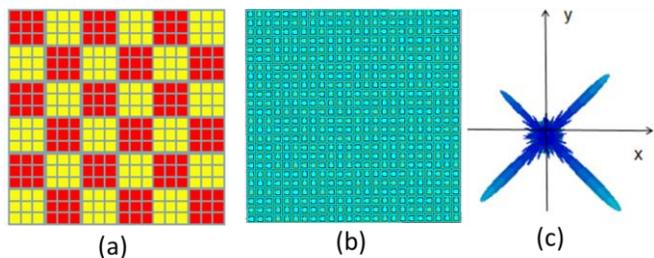

Fig. 10. (a) The schematic of the checkerboard encoding metasurface sequence, (b) The structure of the checkerboard coding pattern, and (c) The corresponding scattering pattern.

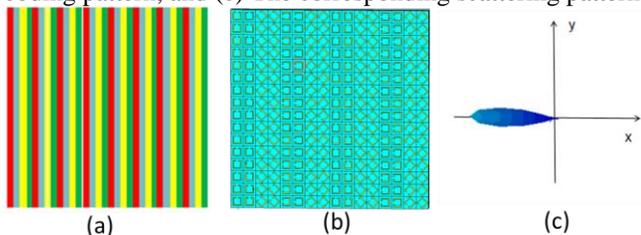

Fig. 11. (a) The schematic of the gradient encoding metasurface sequence, (b) The structure of encoding metasurface, and (c) The corresponding scattering pattern.

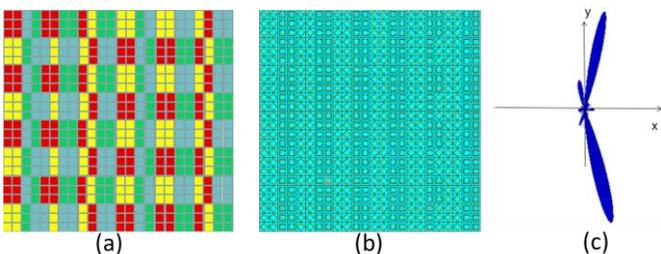

Fig. 12. (a) The schematic of a new checkerboard encoding metasurface sequence after the addition operation, (b) the corresponding structure of an encoding metasurface, and (c) the corresponding scattering pattern.

## V. Conclusions

To improve the transmission efficiency of devices in the terahertz band, we proposed a multi-layer split-ring resonators to develop the coded metasurface. As per the PB phase theory, the phase arrangement of the coded metasurface from 0 °to 360 ° can be realized by simply rotating the unit structure. To achieve different scattering angles, the Fourier convolution principle with the addition and subtraction operations on two different encoding metasurface sequences can be performed.

## References

[1] S. Teng, Q. Zhang, H. Wang, L. Liu, H. Lv, Conversion between polarization states based on metasurface. Photonics Res. **7** (3), 246-250 (2019).

[2] X. Luo, Z. Tan, C. Wang, J. Cao, A reflecting-type highly efficient terahertz cross-polarization converter based on metamaterials. Chin. Opt. Lett. 17(9), 093101- (2019).

[3] L. Koirala, C. Park, S. Lee, D. Choi, Angle tolerant transmissive color filters exploiting metasurface incorporating hydrogenated amorphous silicon nanopillars. Chin. Opt. Lett. **17**(8), 082301- (2019).

[4] T. Hou, Y. An, Q. Chang, P. Ma, J. Li, D. Zhi, L. Huang, R. Su, J. Wu, Y. Ma, P. Zhou, Deep-learning-based phase control method for tiled aperture coherent beam combining systems. High Power Laser Sci. Eng. **7**(4), e59, 2019.

[5] H. Chen, J. Wang, H. Ma, S. Qu, Z. Xu, A. Zhang, M. Yan, Y. Li, Ultra-wideband polarization conversion metasurfaces based on multiple plasmon resonances. J. Appl. Phys. **115**, 154504 (2014).

[6] H. Li, G. Wang, H. Xu, T. Cai, J. Liang, X-band phase-gradient metasurface for high-gain lens antenna application. IEEE Trans. Antenn. Propag. **63**(11), 5144–5149 (2015).

[7] L. Huang, S. Zhang, T. Zentgraf, Metasurface holography: from fundamentals to applications. Nanophotonics. 7(6), 1169-1190(2018).

[8] A. Minovich, A. Miroshnichenko, A. Bykov, T. Murzina, D. Neshev, Y. Kivshar, Functional and nonlinear optical metasurfaces. Laser Photonics Rev. 9(2), 195-213(2015).

[9] H. Wang, J. Zheng, Y. Fu, C. Wang, X. Huang, Z. Ye, L. Qian, Multichannel high extinction ratio polarized beam splitters based on metasurfaces. Chin. Opt. Lett. **17**(5), 052303- (2019).

[10] M. Huault, D. De Luis, J. Apinaniz, M. De Marco, C. Salgado, N. Gordillo, C. Neira, JA. Perez-Hernandez, R. Fedosejevs, G. Gatti, L. Roso, L. Volpe. A 2D scintillator-based proton detector for high repetition rate experiments. High Power Laser Sci. and Eng. **7**(4), 60 (2019).

[11] M. Akram, M. Mehmood, X. Bai, R. Jin, M. Premaratne, W. Zhu, High Efficiency Ultrathin Transmissive Metasurfaces. Adv. Opt. Mater. 7(11), 1801628(2019).

[12] M. Akram, G. Ding, K. Chen, Y. Feng, W. Zhu, Ultrathin Single Layer Metasurfaces with Ultra‐Wideband Operation for Both Transmission and Reflection. Adv. Mater. 32(12), 1907308 (2020).

[13] M. Akram, X. Bai, R. Jin, G. Vandenbosch, M. Premaratne, W. Zhu, Photon Spin Hall Effect-Based Ultra-Thin Transmissive Metasurface for Efficient Generation of OAM Waves. IEEE Trans. Antenn. Propag. 67(7), 4650-4658(2019).

[14] X. Bie, X. Jing, Z. Hong, C. Li, Flexible control of transmitting terahertz beams based on multilayer encoding metasurfaces. Appl. Opt. 57(30), 9070-9077(2018).

[15] Z. Ma, S. Hanham, P. Albella, B. Ng, H. Lu, Y. Gong, S. Maier, M. Hong. Terahertz All-Dielectric Magnetic Mirror Metasurfaces. ACS Photon. 3(6),1010-1018(2016).

[16] T. Cui, M. Qi, X. Wan, J. Zhao, Q. Cheng, Coding metamaterials, digital metamaterials and programmable metamaterials. Light: Sci. Appl. **3**(10), e218(2014).

[17] F. Aieta, P. Genevet, M. Kats, N. Yu, R. Blanchard, Z. Gaburro, F. Capasso, Aberration-free ultrathin flat lenses and axicons at tele-com wavelengths based on plasmonic metasurfaces. Nano Lett. **12**(9), 4932–4936 (2012).

[18] X. Zang, Y. Zhu, C. Mao, W. Xu, H. Ding, J. Xie, Q. Cheng, L. Chen, Y. Peng, Q. Hu, M. Gu, S. Zhuang, Manipulating Terahertz Plasmonic Vortex Based on Geometric and Dynamic Phase. Adv. Opt. Mater. **7**(3), 1801328 (2019).

[19] Q.-T. Li, F. Dong, B.Wang, F. Gan, J. Chen, Z. Song, L. Xu, W. Chu, Y.-F. Xiao, Q. Gong, and Y. Li, Polarization independent and high-efficiency dielectric metasurfaces for visible light, Opt. Express 24, 16309 (2016).

[20] A. Arbabi, A. Faraon, Fundamental limits of ultrathin metasurfaces, Sci. Rep. 2017, 7, 43722.

[21] A. Forouzmand, S. Tao, S. Jafar-Zanjani, J. Cheng, M. M. Salary, H. Mosallaei, Double split-loop resonators as building blocks of

metasurfaces for light manipulation: bending, focusing, and flat-top generation, J. Opt. Soc. Am. B 2016, 33, 1411.

[22] D. Zhang, X. Yang, P. Su, J. Luo, H. Chen, J. Yuan, L. Li, J. Phys. D: Appl. Phys. 2017, 50, 495104.

[23] Paquay M, Iriarte JC, Ederra I, Gonzalo R, de Maagt P. Thin AMC structure for radar cross-section reduction. IEEE Trans Antennas Propag 2007; 55: 3630–3638.

[24] C. E. Garcia-Ortiz, R. Cortes, J. E. Gómez-Correa, E. Pisano, J. Fiutowski, D. A. Garcia-Ortiz, V. Ruiz-Cortes, H.-G. Rubahn, and V. Coello, "Plasmonic metasurface Luneburg lens," Photon. Res. **7**, 1112-1118 (2019).

[25] Chawin Sitawarin, Weiliang Jin, Zin Lin, and Alejandro W. Rodriguez, "Inverse-designed photonic fibers and metasurfaces for nonlinear frequency conversion [Invited]: publisher's note," Photon. Res. **7**, 493-493 (2019).

[26] Bao Du, Hong-Bo Cai, Wen-Shuai Zhang, Shi-Yang Zou, Jing Chen, and Shao-Ping Zhu, "A demonstration of extracting the strength and wavelength of the magnetic field generated by the Weibel instability from proton radiography," High Power Laser Science and Engineering 7(3), e40 (2019).

[27] Shimon Rubin,Yeshaiahu Fainman. "Nonlinear, tunable, and active optical metasurface with liquid film," Advanced Photonics, 2019, 1(6): 066003.

[28] He, X.; Lin, F.; Liu, F.; Zhang, H. Investigation of Phonon Scattering on the Tunable Mechanisms of Terahertz Graphene Metamaterials. *Nanomaterials 10*, 39, 2020.

[29] Xiaoyong He, Fangting Lin, Feng Liu, and Wangzhou Shi, Tunable strontium titanate terahertz all-dielectric metamaterials, Journal of Physics D: Applied Physics, 53(15), 155105, 2020.

[30] Jun Peng, Xiaoyong He, Chenyuyi Shi, Jin Leng, Fanting Lin, Feng Liu, Hao Zhang, Wangzhou Shi, Investigation of graphene supported terahertz elliptical metamaterials, Physica E, 124, 114309, 2020.

[31] Han Wang, Lixia Liu, Changda Zhou, Jilian Xu, Meina Zhang, Shuyun Teng, Yangjian Cai, Vortex beam generation with variable topological charge based on a spiral slit, Nanophotonics 8 (2) (2019) 317–324.

[32] Wang Qizhang Han, Lixia Liu, Shuyun Teng, Generation of vector beams using spatial variation nanoslits with linearly polarized light illumination, Opt. Express 26 (2018) 24145–24154.

[33] Han Wang, Lixia Liu, Chunxiang Liu, Xing Li, Shuyun Wang, Qing Xu, Shuyun Teng, Plasmonic vortex generator without polarization dependence, New J. Phys. 20 (2018) 033024.

[34] Lixia Liu, Han Wang, Yuansheng Han, Xiaoqing Lu, Haoran Lv, Shuyun Teng, Color filtering and displaying based on hole array, Opt. Commun. 436, 96–100 (2019).

[35] M. King, N. M. H. Butler, R. Wilson, R. Capdessus, R. J. Gray, H. W. Powell, R. J. Dance, H. Padda, B. Gonzalez-Izquierdo, D. R. Rusby, N. P. Dover, G. S. Hicks, O. C. Ettlinger, C. Scullion, D. C. Carroll, Z. Najmudin, M. Borghesi, D. Neely, and P. McKenna, "Role of magnetic field evolution on filamentary structure formation in intense laser–foil interactions," High Power Laser Science and Engineering 7(1), e14 (2019).